\documentclass[]{spie}
\usepackage{amssymb}
\usepackage[]{graphicx}

\title{Active Matter on    
Asymmetric Substrates} 
\author{C.J. Olson Reichhardt, J. Drocco, T. Mai, M.B. Wan, and C. Reichhardt 
\skiplinehalf
Theoretical Division, 
Los Alamos National Laboratory, Los Alamos, New Mexico USA 87545}

\authorinfo{Further author information: 
E-mail: charlesr@cnls.lanl.gov, Telephone 1 505 667 9958}

\begin{document}
\maketitle

\begin{abstract}
For collections of particles in a thermal bath interacting with an asymmetric
substrate, it is possible for a ratchet effect to occur where the particles
undergo a net dc motion in response to an ac forcing.
Ratchet effects have been demonstrated in a 
variety of systems including colloids as well as
magnetic vortices in type-II superconductors.
Here we examine the case of active matter 
or self-driven particles interacting with asymmetric substrates. 
Active matter systems include  
self-motile colloidal particles undergoing catalysis, swimming
bacteria, artificial swimmers, crawling cells, 
and motor proteins. 
We show that
a ratchet effect can arise in this 
type of system even in the absence of ac forcing.
The directed motion occurs for certain particle-substrate interaction
rules and its magnitude depends on the amount of
time the particles spend swimming in one direction before turning and
swimming in a new direction.
For strictly Brownian particles there is no ratchet effect. If the
particles reflect off the barriers or 
scatter from the barriers according to Snell's law there is no
ratchet effect; 
however, if the particles can align with the barriers or move along the
barriers, directed motion arises. We also find that under certain 
motion rules,
particles accumulate along the walls of the container
in agreement with experiment. We
also examine pattern formation for synchronized particle motion. 
We discuss possible applications of this system for 
self-assembly, extracting work, 
and sorting as well as future directions such as considering
collective interactions and flocking models. 
\end{abstract}

\keywords{Colloid, optical traps, active matter}

\section{INTRODUCTION}

There have been a number of experiments 
examining the  behavior of colloids interacting with ordered substrates such 
as two-dimensional square and triangular trap arrays 
created by optical means 
\cite{Reichhardt,Bechinger,Trizac,Olson,Frey,Mangold,Roth,Libal}. 
Many of the observed orderings for repulsively interacting colloids
on periodic substrates 
have similar structures
to those of vortices in type-II superconductors with periodic arrays of pinning 
sites \cite{Harada,Commensurate,Jensen,Scalettar}.  
The colloids and vortices can be driven over the substrates 
via an applied current, 
fluid flow, electric fields, or 
other optical means 
\cite{Korda,Korda2,Spalding,Smith,MP,Sm,Sancho,Xiao,Sohn,Speer}. 
It is also possible to create systems with asymmetric substrates
which produce directed motion of particles
even in the absence of an external dc drive \cite{Prost,L}. 
By applying an ac drive or by flashing the substrate,
a net dc flow of particles can be induced 
which is termed a ratchet effect \cite{P}. 
In two dimensions ratchet effects 
can occur in the presence of symmetric substrates when some other
symmetry is broken, such as with a rotating ac drive
\cite{Ratchet,Ratchet2,Tierno} 
or by dynamically switching the 
substrate in a certain order \cite{Hastings,Babic,Lee}.  
The ratchet effect has been demonstrated for colloids on asymmetric substrates
\cite{Prost,L} 
and symmetric substrates with asymmetric drives 
\cite{Tierno,Lee}, for cold atom systems with flashing
substrates \cite{Ren,Lucas}, 
for vortices in type-II superconductors with asymmetric pinning sites 
\cite{Janko,Marchesoni,Olson2,Silva,Lu}, 
for vibrated granular matter on asymmetric substrates 
\cite{Farkas,Wambaugh}, and for
driven liquid drops on asymmetric heated plates \cite{Linke}. 
Applications of the ratchet effect 
include sorting of different particles which ratchet at different rates
over the substrate.
In the overdamped single particle limit the ratchet
effect generally occurs in a single direction; however,
in collections of interacting particles,
ratchet reversals can arise where the dc flow reverses as a parameter
is varied
\cite{Andras,Silva,Lu,Wambaugh}.

In all the ratchet systems mentioned above there is some form of external drive,
such as a single ac drive, flashing of the substrate, 
or multiple external oscillating drives.
There are many systems, termed active matter systems, that exhibit motion 
in the absence of any external drive.
Examples include self-catalyzing colloidal particles \cite{Howse},  
microscopic artificial swimmers \cite{Dreyfus},
swimming bacteria, and active cells \cite{Berg}. 
Recently, experiments were conducted for swimming bacteria
in the presence of an array of asymmetric funnels \cite{Austin}. 
No directed motion occurs for Brownian particles or nonswimming bacteria,
but when swimming bacteria are added, a ratchet effect occurs and the
bacteria concentrate on one side of the funnel array.
Due to the complexity of the bacterial system,
there are a wide variety of possible mechanisms 
for the concentration including hydrodynamic interactions, 
chemotaxis, or collective motions of the bacteria;
however, a simple model of individual particles undergoing run and tumble
dynamics along with a wall-following rule 
reproduces all of
the experimental observations including 
the buildup of particles in the funnel tips 
and along the walls \cite{Wan}.
In this model, the particles move in a single direction
over a run length $l_{r}$ during a fixed run time $\tau_{r}$,
which is the time between tumbling events. 
Every $\tau_{r}$ simulation time steps, the particles
randomly reorient and move in a new single direction 
over a distance $l_{r}$ until the next tumbling event
a time $\tau_r$ later. 
When the particle encounters a barrier, 
it moves along the barrier with a speed determined by the component of
its running motion that is parallel to the barrier
until the next tumbling event or
until reaching the end of the barrier.
If $\tau_{r}$ is small, $l_{r}$ is small and the particles 
act more Brownian-like, with no rectification occurring for
the limit of infinitesimal $\tau_{r}$.
The addition of steric interactions or thermal fluctuations
reduces the rectification, while the rectification is enhanced when
$l_r$ is increased.
With a similar
model it was numerically 
shown and demonstrated in experiment that an asymmetric fly-wheel 
rotates in a preferred direction 
in the presence of a bacterial bath \cite{Ruocco,Ruocco2,Aranson}. 
Other theoretical and numerical work \cite{Tailleur}
has demonstrated rectification 
in swimming bacteria that obey run-and-tumble dynamics under conditions in 
which, due to long running lengths, the particles accumulate and run along the 
walls. This work assumed no hydrodynamic interactions indicating that 
hydrodynamic effects are not required to produce rectification. 
The authors also mention that for bacteria that collide 
elastically with the walls, no rectification occurs.
Other experiments have also shown
directed cellular motion over asymmetric substrates 
\cite{Peter,Shear,Whitesides} and even different directions of motion
for different cell types on the same substrate \cite{Mahmud}. 
Unlike the bacterial case, the mechanics of how the 
eukaryotic cells move is very important and the latter cannot be modeled simply
by point particles. 
These studies open a new field of 
controlling active matter using periodic and asymmetric substrates. 
If the motion 
of self driven particles can be well controlled, 
it may be possible to use the particles to perform 
microscopic work such as transporting larger obstacles or cargo. 
Additionally, biological sorting could be achieved via ratchet effects.
In this work we investigate the rules for run-and-tumble 
particles which generate a net dc motion of active matter 
in the presence of asymmetric barriers. 
We find that if the particles are fully aligned
with the barriers, a much stronger rectification effect occurs. 
If the particles
reflect or scatter off the barriers than no ratchet effect 
occurs even for very large $l_{r}$. 
This shows that the rectification 
requires the breaking of detailed balance in the interactions
with the asymmetric barriers.
We also find that there is a build up of particles along the barriers
similar to what is observed in experiment \cite{Wan}. 
This  build up is due to the
wall following rule combined with the finite size of the system. 
If the tumbling times of the particles 
are synchronized, we find that patterned moving density fronts of particles
occur after the particles accumulate in corners of the containers or in the
funnel tips.

\section{COMPUTATIONAL MODEL}

Figure~1 shows an image of our system which consists of an 
$L \times L$ two-dimensional box
of length $L=99$ 
with confining walls in the $x$ and $y$ directions
containing $N_a$ active particles. In the center
of the system is an array of asymmetric funnels, placed in 
the same geometry
used in the experiments for the swimming bacteria \cite{Austin}. 
The particle density is $\rho = N_{a}/L^{2}$.
The particles move in an overdamped media without hydrodynamic interactions
according to the overdamped equation of motion
\begin{equation}
\eta \frac{d{\bf R}_i}{dt} = {\bf F}^{m}_{i}(t) 
+ {\bf F}_i^{B} + {\bf F}_i^{pp}  
+ {\bf F}^{T}_i.     
\end{equation} 
Here $\eta=1$ is a phenomenological damping term.
The time dependent motor force
${\bf F}^{m}_{i}(t)$ 
is represented by run and tumble dynamics. 
Here  $|{\bf F}^{m}| = 2.0$ while ${\bf \hat F}^{m}_i$
is selected randomly and changed every $\tau_r$ simulation
time steps.
The resulting run length $l_{r} = \tau_{r}\delta t |F^{m}|$ 
where $\delta t = 0.005$ is the simulation time step size. 
Fig.~2(a) shows a single active particle  with $\tau_{r} = 1000$
and $l_{r} = 10$ moving in straight paths for extended
distances before randomly orienting and moving in other directions, producing
a time average velocity of zero.
With smaller
$\tau_{r}$ the length scale over which the motion appears Brownian decreases,
as shown in Fig.~2(b) for $l_{r} = 1.0$. 
Many active matter systems contain thermal fluctuations which we represent
as Langevin kicks with the term
${\bf F}^{T}_{i}$, where
$\langle F^T_i\rangle =0$ and
$\langle F^{T}_{i}(t)F^{T}_{j}(t^{\prime})\rangle 
= 2\eta k_{B}T\delta_{ij}\delta(t - t^\prime)$   
where $k_{B}$ is the Boltzmann constant. 
Fig.~2(c) illustrates the combination of run and tumble dynamics with 
thermal fluctuations.
We initially consider $F^{T} = 0.0$ 
since in many active matter systems at room temperature, such as
bacteria, the thermal fluctuations are small on the scale of the particles.
The term ${\bf F}_{i}^{pp}$ 
is the particle-particle interaction force which is modeled as 
simple short ranged steric repulsion. 
The term ${\bf F}_{i}^{B}$ is the force from the barriers which are 
each modeled
by two half-parabolic domes of strength $F_{B} = 30$ 
and radius of $r_{B} = 0.05$ which repel 
particles along the direction perpendicular to the trap axis.
\begin{equation}
{\bf F^{B}}_{i} = \sum^{N_{B}}_{k = 0}\frac{F_{B}r_{1}}{r_{B}}\Theta(r_{1}){\hat {\bf R}}_{ik}^{\pm}
+ \frac{F_{B}r_{2}}{r_{B}}\Theta(r_{2}){\hat {\bf R}_{ik}}^{\perp} .
\end{equation} 
Here the total number of barriers including the confining walls is $N_{B} = 28$,
$r_{1} = r_{B}-R_{ik}^{\pm}$, $r_{2} = r_{B} - R^{\perp}_{ik}$,  
$R_{ik}^\pm = | {\bf R}_{i} - {\bf R}^{B}_{k} \pm L_{B}{\hat {\bf p}}_{||}^{k}|$, 
${\hat {\bf R}}_{ik}^\pm = ({\bf R}_{i} - {\bf R}^{B}_{k} \pm L_{B} {\hat {\bf p}}_{||}^{k})/R_{ik}^\pm$, 
$R_{ik}^\perp=|({\bf R}_i-{\bf R}_k^B) \cdot {\bf \hat p}_{\perp}^{k}|$,
${\hat {\bf R}}_{ik}^{\perp} = [({\bf R}_{i} - {\bf R}^{B}_{k}) \cdot {\hat {\bf p}}_{\perp}^{k}]/R_{ik}^{\perp}$.
${\bf R}_{i}({\bf R}_{k}^{B})$ is the position of particle $i$ (barrier $k$),
and ${\hat {\bf p}}^{k}_{||}$ (${\hat {\bf p}}^{k}_{\perp}$) 
is a unit vector parallel (perpendicular)
to the axis of barrier $k$.  The funnels are modeled as two barriers meeting at 
a common end point at angles $\theta$ and $\pi - \theta$ with the $x$-axis
and the tips of adjacent funnels are spaced a 
distance $8.25$ apart along the $x$ 
direction. 

\begin{figure}
\begin{center}
\begin{tabular}{c}
\includegraphics[width=5.0in]{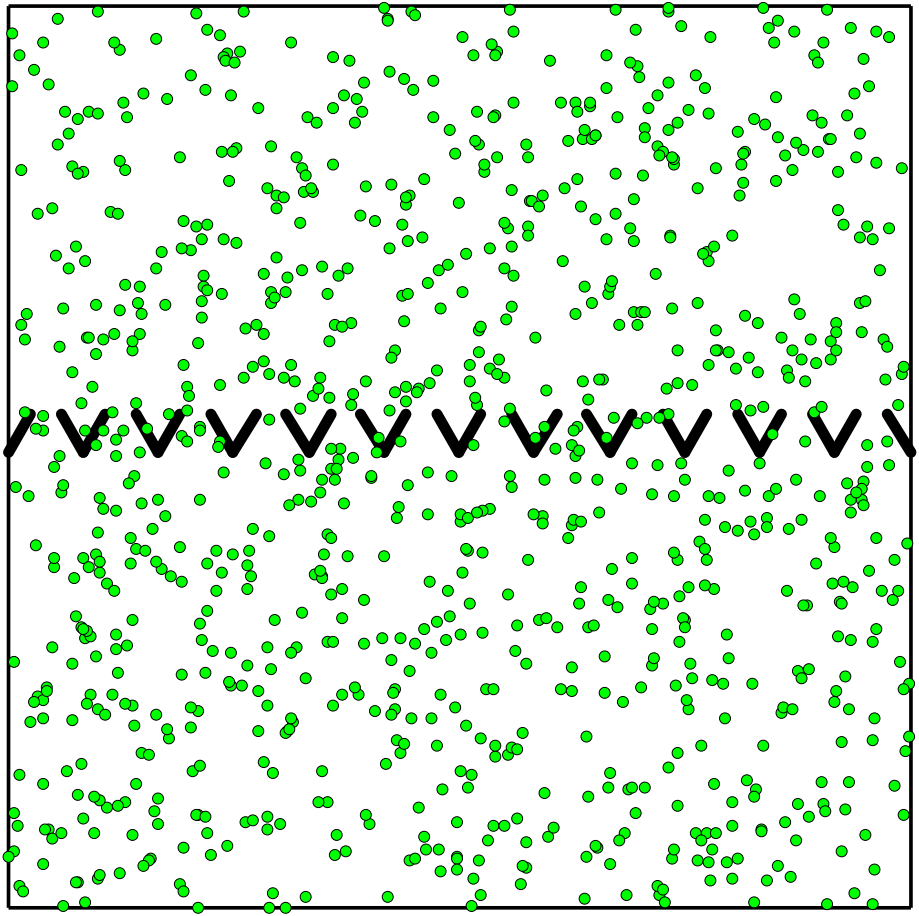}
\end{tabular}
\end{center}
\caption{
A snapshot of the system which consists of a confining 
box with an array of asymmetric barriers (heavy lines)
and $N_{a}$ active particles (green dots) that obey different rules for
motion and interactions with the walls and barriers. 
}
\end{figure}

\section{Rules Producing Rectification for Non-Interacting Particles }

\begin{figure}
\begin{center}
\begin{tabular}{c}
\includegraphics[width=5.0in]{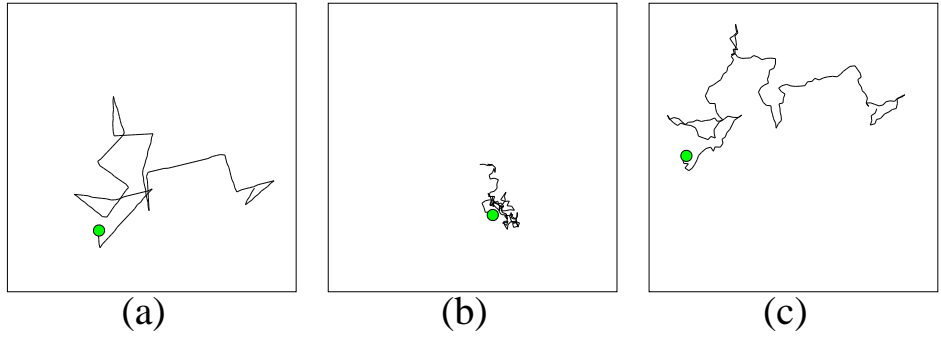}
\end{tabular}
\end{center}
\caption{
The trajectories of a single active particle undergoing run and tumble 
dynamics.
(a) $l_r=10.$
(b) $l_{r} = 1$. 
(c) The same as in (a) 
but with an added thermal noise component with $F^{T} = 10.0$. 
}
\end{figure}

We first consider the case of no thermal fluctuations and no steric
interactions between particles.
In previous work, a particle encountering a barrier had the component of its
motion perpendicular to the barrier canceled but experienced no
reorientation of ${\bf F}_i^m$ by the barrier \cite{Wan}.
This behavior, termed Rule I, is is illustrated in the schematic 
in Fig.~3(a). 
Here we consider three additional barrier interactions.
In Rule II, ${\bf F}_i^m$ is realigned to be parallel with the barrier.
This is a realistic assumption
for many types of active matter such as elongated bacteria, which can
align their swimming with a wall due to hydrodynamic interactions.
In Rule III, ${\bf F}_i^m$ is reversed so that the particles reflect back into
the direction from which they came, as shown in Fig.~3(c), while
in Rule IV, ${\bf F}_i^m$ is reflected across a line perpendicular to
the barrier so that the particles scatter off the barrier, as illustrated
in Fig.~3(d).

\begin{figure}
\begin{center}
\begin{tabular}{c}
\includegraphics[width=5.0in]{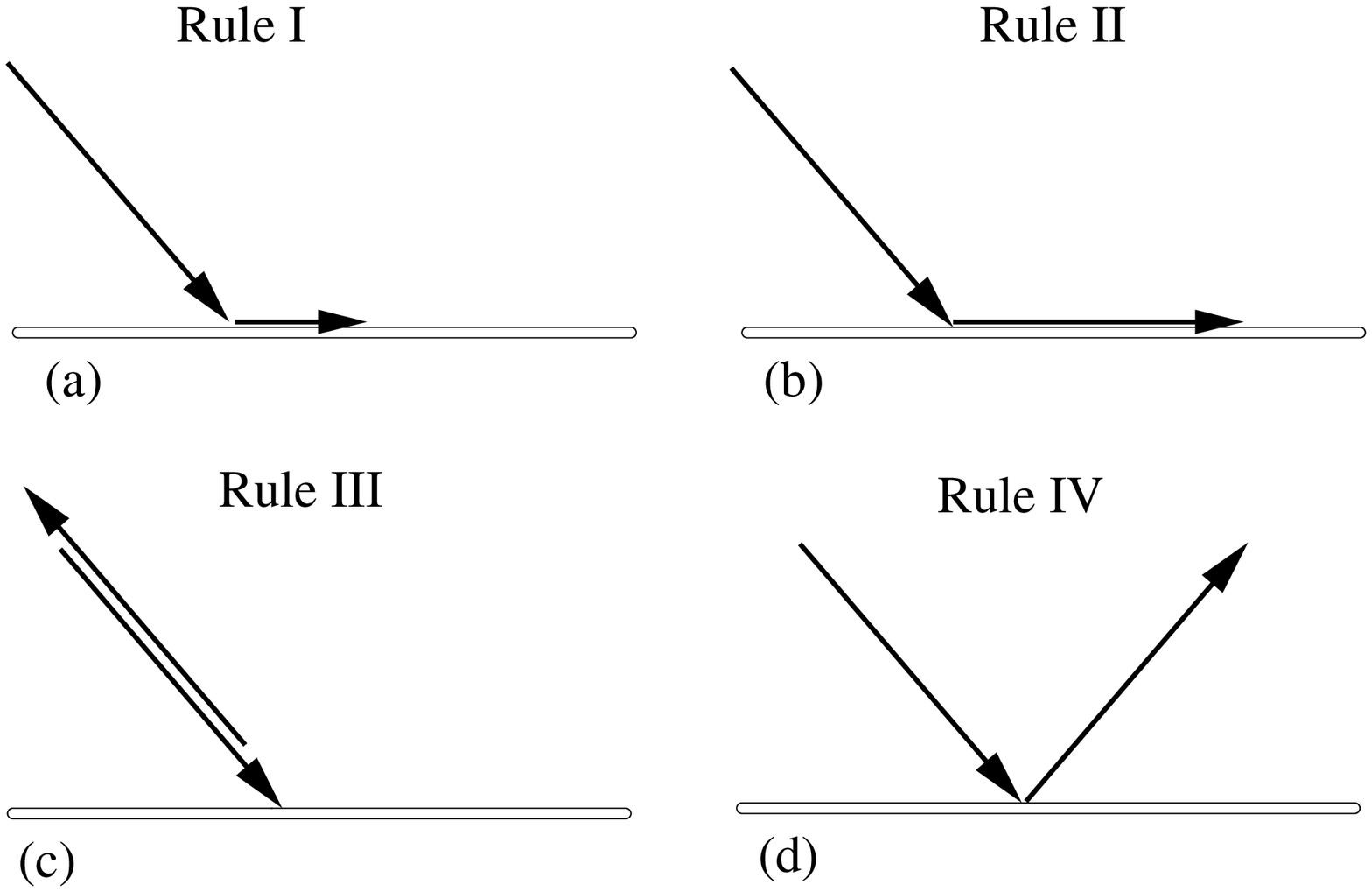}
\end{tabular}
\end{center}
\caption{
Different rules for the particle-barrier interactions. 
(a) Rule I. The particles
move with the component of ${\bf F}_i^m$ 
that is parallel to the barrier.
(b) Rule II. ${\bf F}_i^m$ is realigned to be completely
parallel to the barrier.
(c) Rule III.  The particles reflect off the barriers. 
(d) Rule IV.  The particles scatter off the barriers.
}
\end{figure}

\begin{figure}
\begin{center}
\begin{tabular}{c}
\includegraphics[width=5.0in]{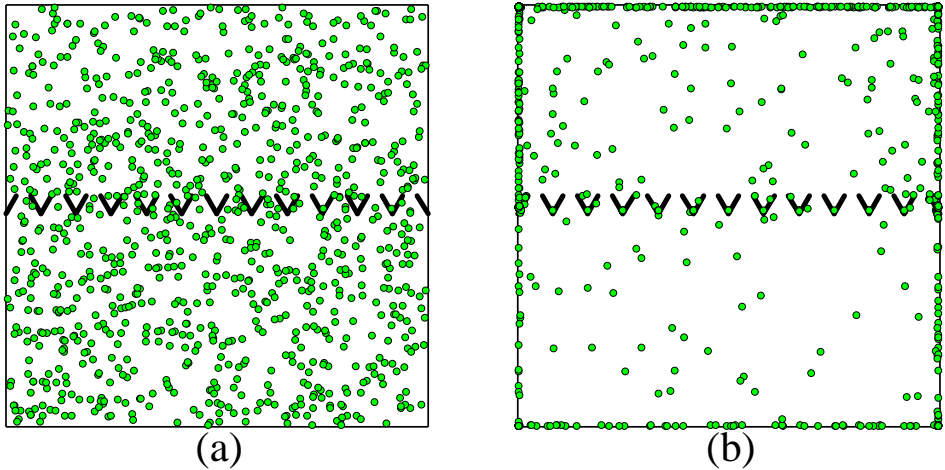}
\end{tabular}
\end{center}
\caption{ 
Image of the sample after 100 tumbling events for 
Rule I. (a) At $l_{r} = 0.01$, the particle density is equal on both sides
of the chamber. (b) At $l_{r} = 180$ the particle density has
substantially increased in the upper chamber with particles accumulating
along the walls and in the funnel tips. 
}
\end{figure}

We now consider Rule I, partial alignment with the barriers, for 
short and long $l_{r}$ as illustrated in Fig.~4. 
For $l_{r} = 0.01$ after 
100 tumbling events in Fig.~4(a),
there is no rectification and the density $\rho^{(1)}/\rho$ in the upper channel
equals the density $\rho^{(2)}/\rho$ in the lower channel,
as shown in Fig.~5.
For  $l_{r} = 180$, there is a buildup
of particles in the upper chamber as shown in Fig.~4(b), and the
dependence of $\rho^{(1)}/\rho$ and $\rho^{(2)}/\rho$ on time has a shape
very similar to the form found in experiments \cite{Austin}, as illustrated
in Fig.~5. 
The rectification mechanism for long $l_r$ is illustrated in Fig.~6.  When
a particle approaches a funnel from below, it is guided along the barrier and
moves into the upper chamber as long as $\tau_r$ is long enough that it does
not change direction before reaching the upper chamber.  On the other hand, 
when a particle approaches a funnel from above, it is guided into the closed
tip of the funnel where it remains trapped until a new tumbling event sends
it back into the upper chamber.
For small $l_{r}$ the same barrier interaction rules still apply; 
however, due to the short time between tumbling events, the particle spends
so little time moving along the barrier that its motion cannot be guided
and no rectification occurs.
This result shows that it is the combination
of the long running time and the breaking of detailed balance 
in the particle-barrier interactions
that lead to the rectification. 
Fig.~7 shows a plot of $r = \rho^{(1)}/\rho^{(2)}$  versus $l_{r}$
for the Rule I sample.
The amount of rectification increases with increasing 
$l_{r}$.  In the limit of small $l_{r}$ 
where the interactions between the particles and the barriers
are in the Brownian limit there is no rectification. 

\begin{figure}
\begin{center}
\begin{tabular}{c}
\includegraphics[width=5.0in]{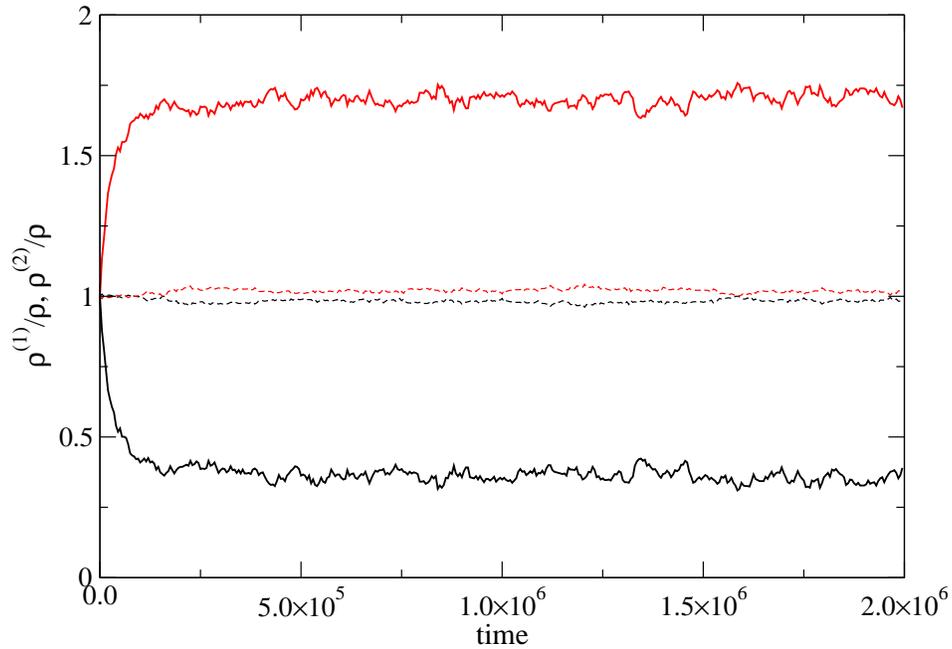}
\end{tabular}
\end{center}
\caption{
$\rho^{(1)}/\rho$ and $\rho^{(2)}/\rho$ vs time in simulation time
steps for the system
in Fig.~4 under Rule I. 
Dotted lines: $l_{r} = 0.01$ from Fig.~4(a) where the density
is almost the same in each chamber. 
Upper heavy line: $\rho^{(1)}/\rho$ for the
$l_{r} = 180$ system from Fig.~4(b); lower heavy line:
$\rho^{(2)}/\rho$ for the system from Fig.~4(b).
}
\end{figure}

\begin{figure}
\begin{center}
\begin{tabular}{c}
\includegraphics[width=5.0in]{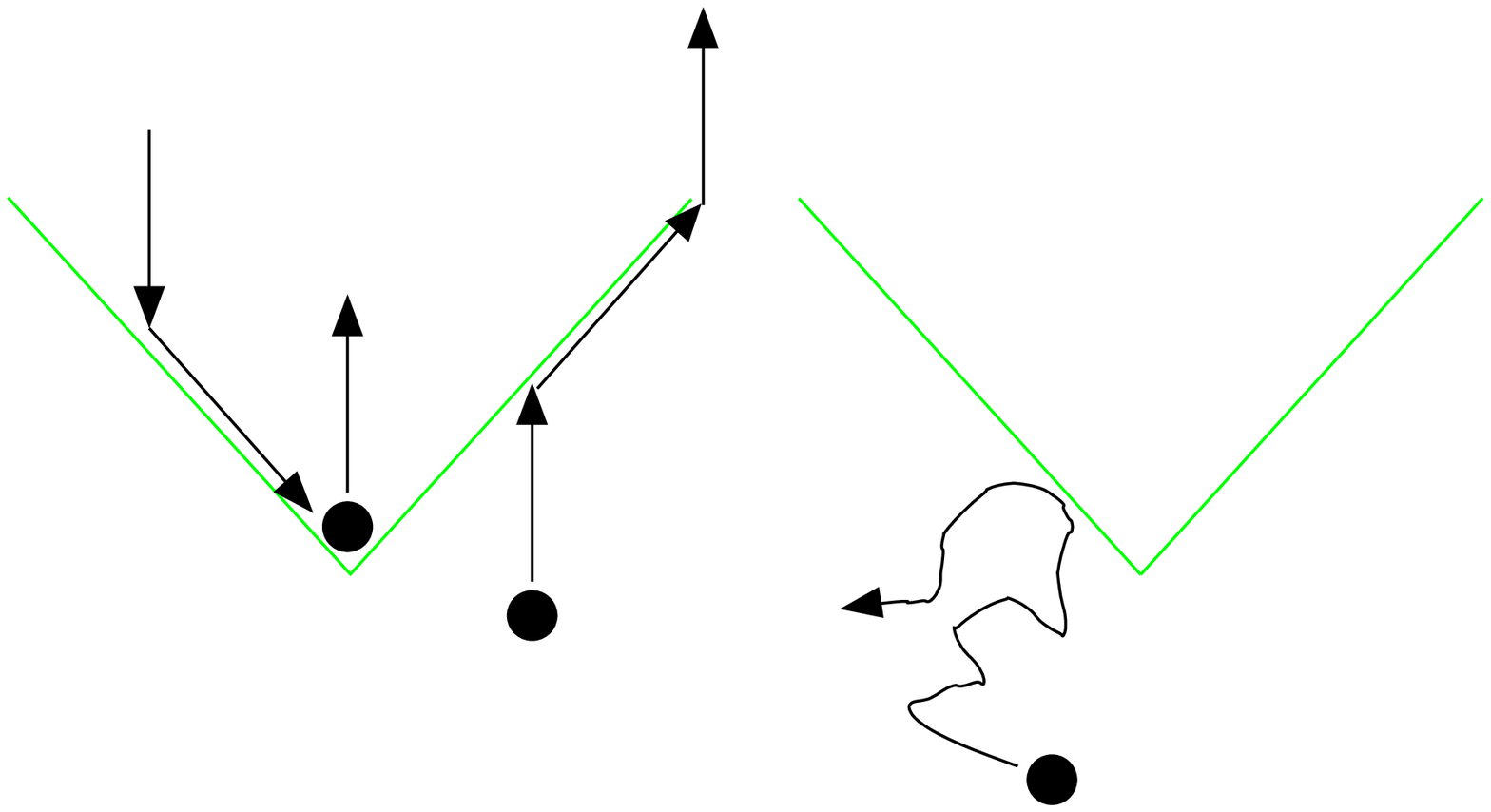}
\end{tabular}
\end{center}
\caption{
A schematic of the rectification mechanism under Rule I and Rule II. 
Left panel: A particle approaching the funnel from above aligns with
the barrier and moves into the funnel tip where it is trapped.
After the next tumbling event the particle may escape from the funnel tip
and move back into the upper chamber.  A particle approaching the funnel
from below aligns with the barrier and is guided into the upper chamber.
Right panel: For very short $l_r$, the particle spends little to no time
interacting with the barrier and its motion cannot be guided by the
spatial asymmetry of the barrier.
}
\end{figure}

\begin{figure}
\begin{center}
\begin{tabular}{c}
\includegraphics[width=5.0in]{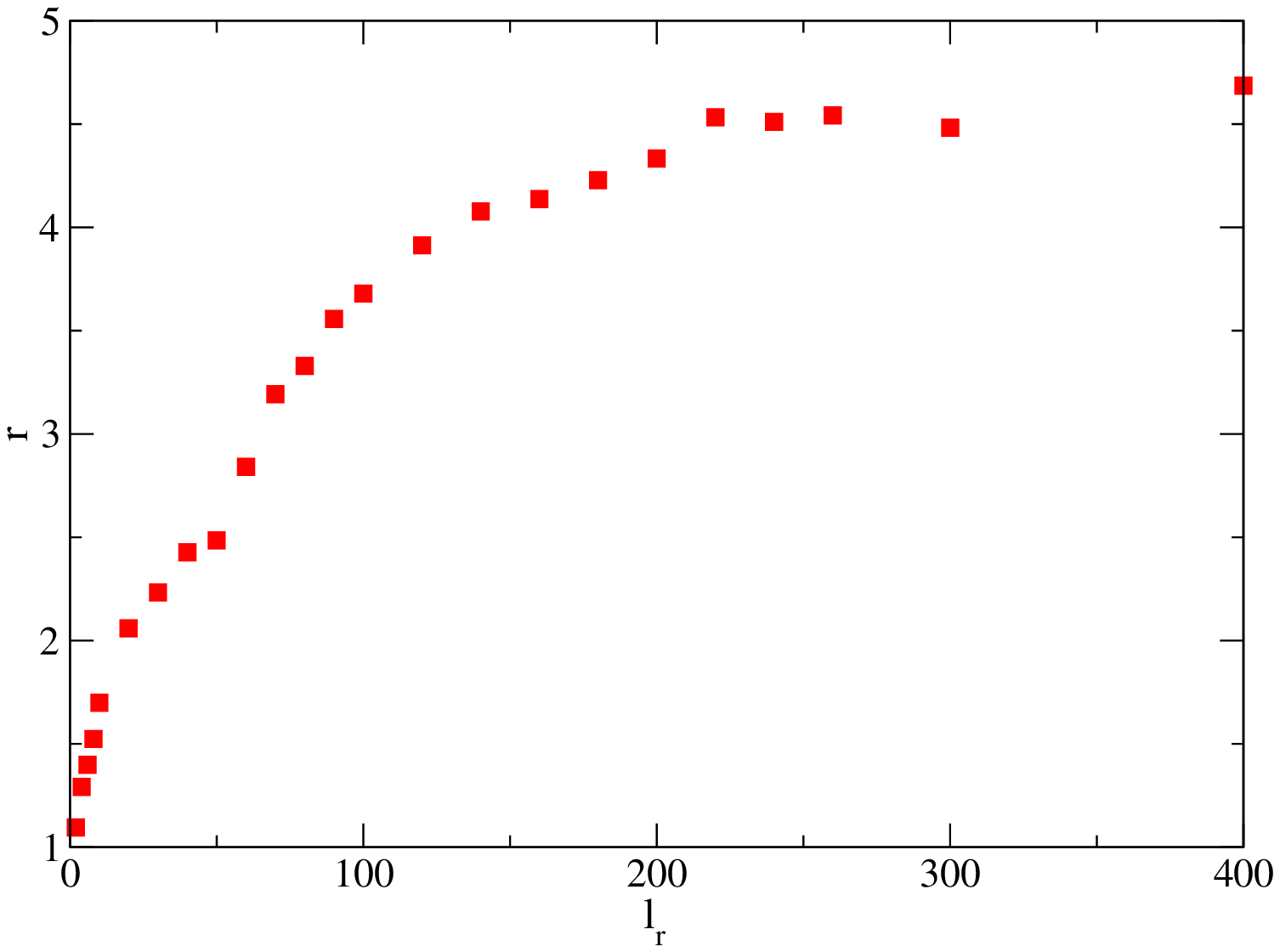}
\end{tabular}
\end{center}
\caption{
The ratio of the particle densities $r = \rho^{(1)}/\rho^{(2)}$ vs $l_{r}$ 
for the system with Rule I. For small $l_{r}$ the rectification is lost. 
}
\end{figure}

When we employ Rule III or Rule IV we find no
rectification, as shown in Fig.~8 
where we plot $r$ for the four different rules in a sample
with $l_{b} = 300$.
Regardless of the value of $l_r$, Rules III and IV do not produce
rectification of the particles due to the fact that they
preserve detailed balance for the interactions of the particles with
the barriers,
which is enough to prevent rectification.  

\begin{figure}
\begin{center}
\begin{tabular}{c}
\includegraphics[width=5.0in]{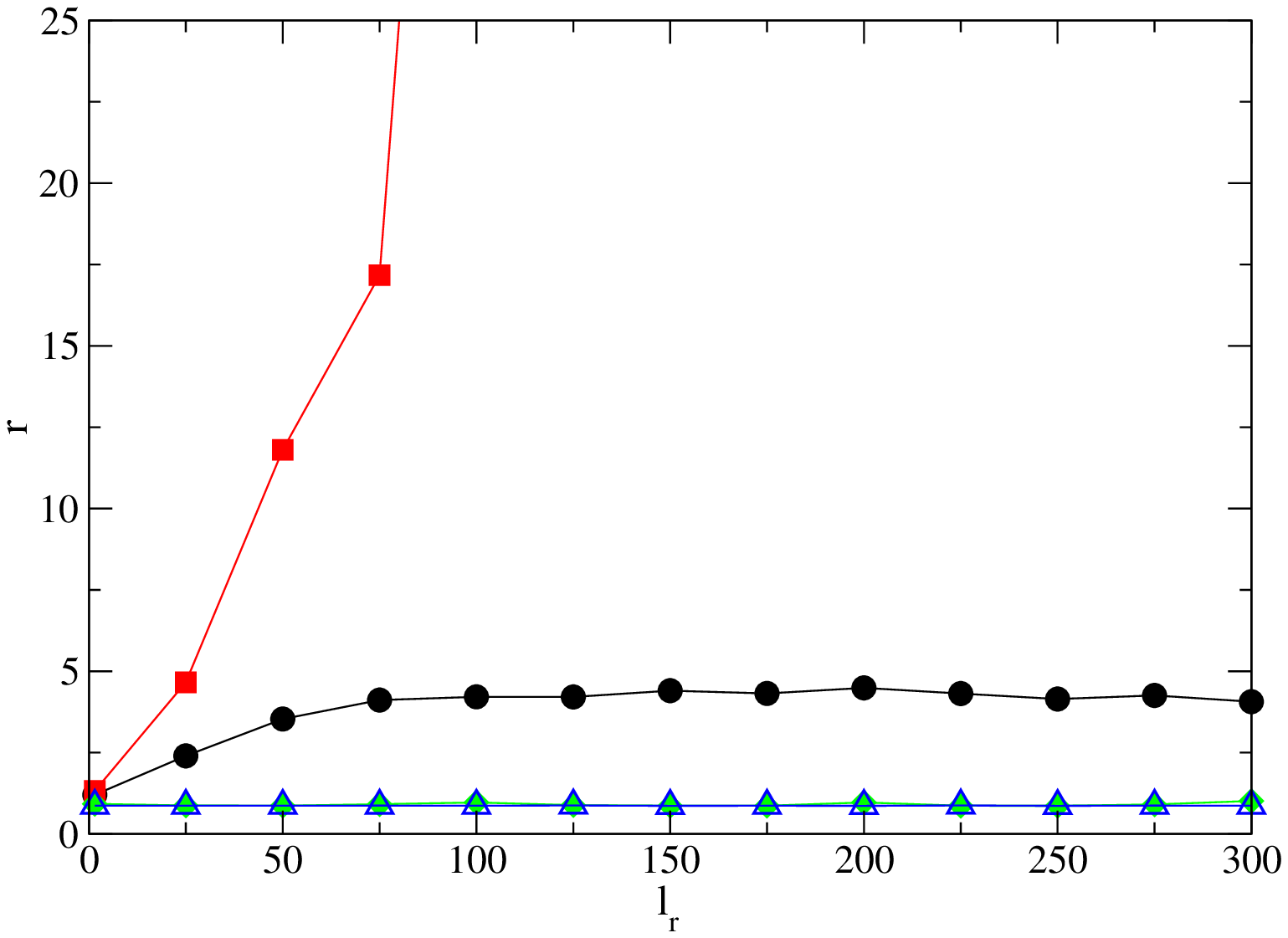}
\end{tabular}
\end{center}
\caption{
The value of $r$ after 6000 tumbling events for a system with $l_{r} = 300$ 
for Rule I (filled circles), Rule II (filled squares), 
Rule III (filled diamonds), and Rule IV (open triangles). Here
only Rules I and II produce rectification.  
}
\end{figure}

We note that the amount of rectification varies 
with other system parameters such as the
angle of the funnels, the overall size of the funnels relative to
$l_b$, and the spacing between adjacent funnels. In general, if the 
funnels are made larger for fixed $l_r$ the rectification is reduced, 
since this has the same effect as holding the funnel size constant and
reducing $l_r$.
Increasing the spacing between the funnels reduces the effect since there is
simply more space for particles to freely pass from the upper chamber to the
lower chamber, producing a larger amount of backflow and lowering the
efficiency of the ratchet.
As a function of the angle between the funnel arms,
the ratchet effect is generally optimized near 50 to 60 
degrees \cite{Wan}.

\section{Steric and Thermal Interactions}

We next consider the effect of adding a short range 
steric repulsion between the particles which is modeled 
as a stiff spring of finite radius which produces a force
\begin{equation}
{\bf F}^{S}_{i} = \sum^{N_{a}}_{i\neq j}\frac{f_{a}r_{3}}{r_{s}}\Theta(r_{3}){\bf \hat R}_{ij}   
\end{equation}
Here $r_{3} = r_{a} - R_{ij}$, $R_{ij} = |{\bf R}_{i} - {\bf R}_{j}|$, 
${\bf \hat R}_{ij} = ({\bf R}_{i} - {\bf R}_{j})/R_{ij}$, and 
$f_{a}=150$ is the force coefficient.

In Fig.~9(a) we plot $r$ vs $\rho$ for a system with $l_{r} = 20$ 
and steric interactions under 
Rule I. 
Here the rectification 
decreases with increasing density.  The interacting particles interfere
with each other's motion in several ways.  Particles that strike the barriers
at an angle that is nearly perpendicular to the barrier move very slowly
along the barrier.  These particles get in the way of other particles
that strike the barrier at lower angles and that would otherwise move
relatively quickly along the barrier.  This interaction decreases the
efficiency of the ratchet.
The buildup of particles along the walls also becomes much more apparent
for the sterically interacting particles, as illustrated in Fig.~10.
A similar formation of excess density along the walls occurred in
the experiments of Ref.~\cite{Austin}.
In Fig.~9(b) we illustrate 
the effects of adding thermal noise to a system of particles without
steric interactions under Rule I with $l_{r} = 20$. 
The plot of $r$ versus $F^{T}$ shows that the rectification is strongly
reduced as the thermal fluctuations begin to dominate the motion of the
particles and the system enters the Brownian limit.    

\begin{figure}
\begin{center}
\begin{tabular}{c}
\includegraphics[width=5.0in]{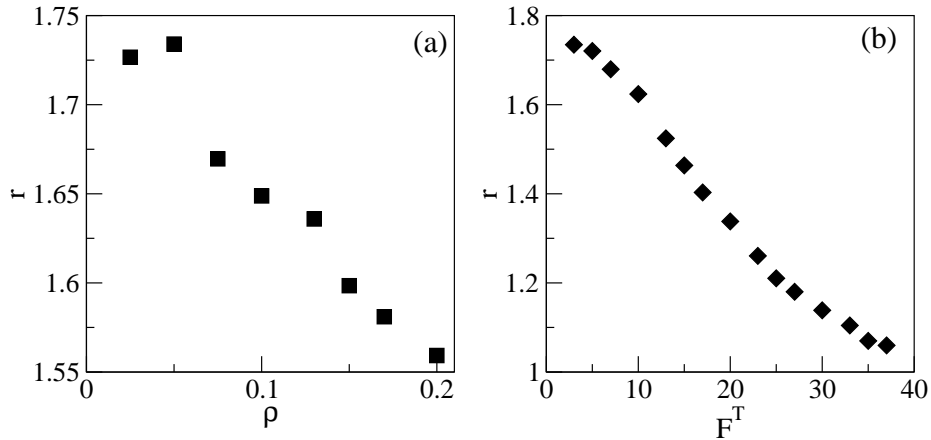}
\end{tabular}
\end{center}
\caption{
(a)  $r$ vs $\rho$ for a system with steric interactions under Rule I 
at $l_{r}  = 20$. Here $r$ decreases with increasing $\rho$. 
(b) $r$ vs the thermal force $F^{T}$ for a system with 
no steric interactions under Rule I at
$l_{r} = 20$.  The rectification is reduced as the magnitude of the
thermal fluctuations increases.
}
\end{figure}

\begin{figure}
\begin{center}
\begin{tabular}{c}
\includegraphics[width=5.0in]{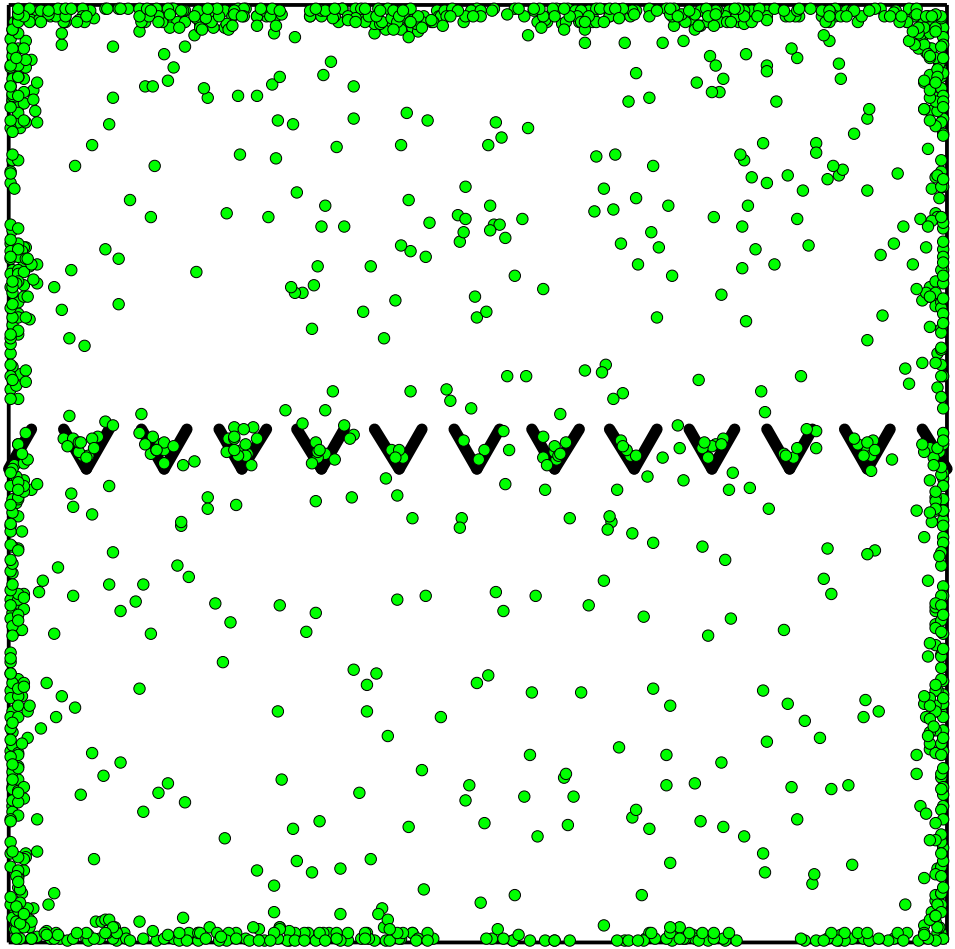}
\end{tabular}
\end{center}
\caption{
The build up of particles along the walls and inside the funnels for a system 
with Rule I and sterically interacting particles at $l_r = 120$. 
This same effect is observed in experiments. 
}
\end{figure}

\section{Synchronized Switching and Pattern Formation} 

Up to this point we have
considered the effects of random tumbling events.  All of the particles
had equal $\tau_r$; however, at the beginning of the simulation
each particle was assigned a randomly chosen
amount of time $0<t<\tau_r$ after which the first tumbling event for that
particle would occur.  This desynchronized the switching of the particles.
It could, however, be possible 
for certain systems to have synchronized tumbling events for all particles.
For example, this would occur if a
periodically applied external field activated the tumbling, so that all
of the particles would change direction simultaneously.
For open systems or samples with periodic boundary
conditions, pattern formation would not be expected to occur; however,
in the confined box and funnel geometry that we study 
we find that pattern formation in the form of expanding fronts of
particles occurs when the switching events are synchronized for particles
that have no steric interactions.
In Fig.~10(a,b) we illustrate the formation of moving fronts 
of particles in a system with no steric interactions at $l_r=80$ under Rule I
with synchronized tumbling times.
Density fronts are emitted by the four corners 
of the container and to a lesser extent also from the ends of the funnels. 
These fronts move out in a growing circular 
pattern, decreasing in density as they spread
until the structures break up due to 
interactions with the walls. 
The pattern forms since the particles can effectively collapse into the
corners and the funnel tips due to their lack of steric interactions; the
particles are trapped in these locations due to the barrier interaction Rule I.
If $\tau_{r}$ is long enough essentially all the
particles will become trapped in the corners and funnel tips. 
Since the tumbling events are synchronized,
the particles move out from each trapping point in a circular pattern after
a switching event. In our system we observe only a semicircle of motion
due to the confinement by the walls.
The fronts are clearly defined only for sufficiently large $\tau_{r}$ since
for short $\tau_{r}$ the particles do not have enough time to 
become highly concentrated in the corners and funnel tips. 
This type of pattern formation is strongly 
reduced for particles with steric interactions since the interactions
reduce the ability of the particles to become concentrated at a point; 
however, some remnant of the moving fronts still can be observed when
steric interactions are present.   

\begin{figure}
\begin{center}
\begin{tabular}{c}
\includegraphics[width=5.0in]{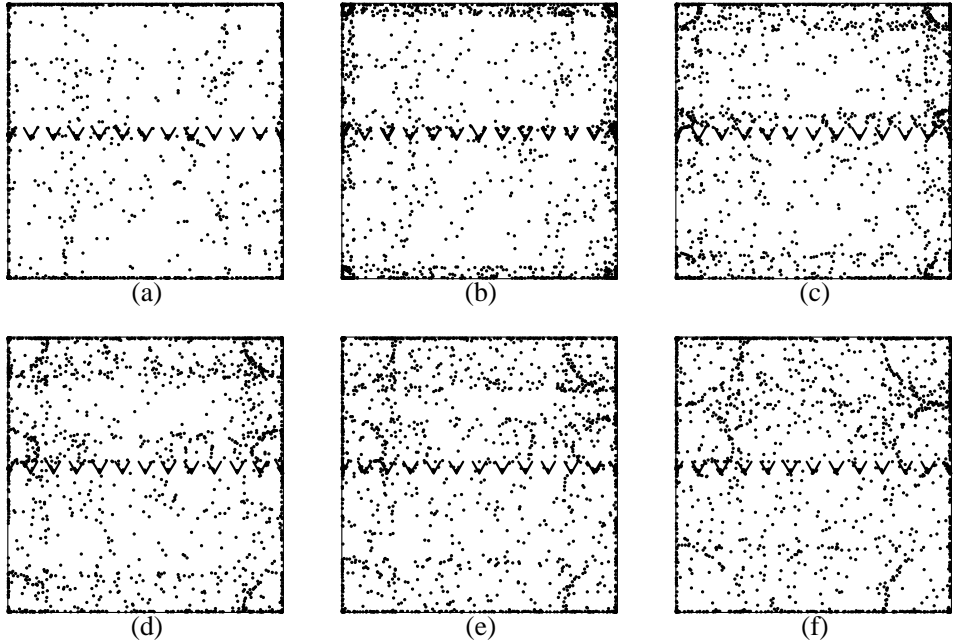}
\end{tabular}
\end{center}
\caption{
The formation of moving particle fronts for a system with Rule I,
no steric interactions, $l_{r} = 80$, and synchronized tumbling times, shown
as consecutive snapshots of the system taken at equal time intervals.
The particles accumulate in the corners of the container and are
released as a semicircular density wave after each switching event.
}
\end{figure}

\section{Discussion and Future Directions} 
This work in conjunction with experiments shows 
that simple substrates can be used
to create directed motion in self-driven particle systems. 
This opens the possibility
for an entirely new class of 
ratchets based on microscale systems 
which could be used
for sorting different species of active particles, 
guiding particles to move in certain directions,
and possibly even extracting work out of such systems. 
Our results show that thermal
fluctuations can strongly reduce the efficiency of the 
directed motion. For large colloids
and many biological systems, thermal effects are generally small on the
scale of the particles;
however, if nanoscale self-driven particles
are considered then thermal effects will be relevant.
Our work and other models of 
this type of system include only point particles or elongated
rods. 
It would be interesting to study ratcheting behavior for
particles with more complicated shapes 
or for particles that contain internal degrees of freedom.
For example, eukaryotic cells exhibit different types 
of mechanical mechanisms for
locomotion such as pushing or pulling motions, and it could be possible to
implement swimming rules that mimic
such motions. 
We have studied only static substrates in our system;
however, it should be possible to create dynamic 
substrates for self-driven particles similar to flashing ratchets,
and such dynamic substrates 
 might be much more effective in directing the motion of the particles. 
It would also be interesting to study different types 
of collective effects such as
by introducing models for flocking or swarming behaviors 
or by adding longer range
interactions between  particles 
which could arise through hydrodynamic effects. 
Although the model we consider is mostly relevant for microscopic systems, 
similar effects could be
tested for larger scale systems such 
as individual or collectively moving insects, animals, and birds. 
Another possibility would be to study self-driven particles 
that have an intrinsic asymmetry as they move through
symmetric or asymmetric substrates.

\section{SUMMARY}
We  have shown that a new type of ratchet 
system can arise in systems of self-driven 
particles or active particles in asymmetric geometries. 
Unlike most ratchet systems where some 
form of external ac force or flashing potential is required to
induce a ratchet effect, the self-driven 
particle system can exhibit a ratchet effect in the absence 
of external forcing.
In our system we consider active particles that 
move with a run-and-tumble
dynamics of the type found in bacterial systems 
where the particles move in a single 
direction for a fixed time before undergoing a tumbling event, randomly
reorienting, and then moving for  
a fixed period of time in a new direction. We studied
different rules for the interactions between the
particles and the barriers, including
alignment or partial alignment with the barrier which causes
the particles to follow the barriers as well as
reflection and scattering interactions. 
When the particles are allowed to move along 
the barriers and when the time between tumbling events is sufficiently long,
we observe rectification of the particles by the barriers. 
We also find a buildup of particles along the walls and 
inside the funnel tips. 
Both
the rectification and the buildup of the particles 
along the barriers are in excellent agreement
with experimental observations. 
Under these same wall interaction rules, if the time between tumbling events
is very small the motion of the
particles becomes more Brownian-like on the length scale of the barriers
and the rectification is lost. 
For reflection and scattering
of the particles off the barriers 
the rectification is lost even for very long times between tumbling events.
These results show that it is the combination of 
the running length and the breaking of detailed
balance in the particle-barrier interactions 
that produce the rectification. We have also
found that if the tumbling times for the particles are synchronized, 
pattern formation occurs in the form of  
moving density fronts generated after each
switching event due to the concentration of particles 
in the corners and funnel tips.
The addition of thermal effects or steric interactions generally 
reduces the 
effectiveness of the rectification. 
Future directions to pursue include more complicated 
shapes and interactions for the individual particles 
as well as the addition of collective effects.   

This work was carried out under the auspices of the NNSA of the
U.S. Department of Energy at Los Alamos National Laboratory
under Contract No. DE-AC52-06NA25396.


\begin{thebibliography}{99}

\bibitem{Reichhardt}
C. Reichhardt and C.J. Olson,
``Novel colloidal crystalline states on two-dimensional periodic substrates.''
{\em Phys. Rev. Lett.} {\bf 89}, pp. 248301, 2002.   

\bibitem{Bechinger}
M. Brunner and C. Bechinger,
``Phase behavior of colloidal molecular crystals on 
triangular light lattices.''
{\em Phys. Rev. Lett.} {\bf 88}, pp. 248302, 2002.   

\bibitem{Trizac}
R. Agra, F. van Wijland, and E. Trizac, 
``Theory of orientational ordering in colloidal molecular crystals.'' 
{\em Phys. Rev. Lett.} {\bf 93}, pp. 018304, 2004.

\bibitem{Olson}
C. Reichhardt and C.J. Olson Reichhardt, 
``Ordering and melting in colloidal molecular crystal mixtures.'' 
{\em Phys.~Rev.~E.} {\bf 71}, pp. 062403, 2005.

\bibitem{Frey} 
A. Sarlah, T. Franosch, and E. Frey, 
``Melting of colloidal molecular crystals on triangular lattices.''
{\em Phys.~Rev.~Lett.} {\bf 95}, pp. 088302, 2005.

\bibitem{Mangold}
K. Mangold, P. Leiderer, and C. Bechinger, 
``Phase transitions of colloidal monolayers in periodic pinning arrays.''
{\em Phys.~Rev.~Lett.} {\bf 90}, pp. 158302, 2003.

\bibitem{Roth}
J.~Mikhael, J.~Roth, L.~Helden, and C.~Bechinger, 
``Archimedean-like tiling on decagonal quasicrystalline surfaces.''
{\em Nature} {\bf 454}, pp. 501--504, 2008.

\bibitem{Libal}
A. Lib{\' a}l, C. Reichhardt, and C.J. Olson Reichhardt, 
``Realizing colloidal artificial ice on arrays of optical traps.''
{\em Phys.~Rev.~Lett.} {\bf 97}, pp. 228302, 2006.

\bibitem{Harada}
K. Harada, O. Kamimura, H. Kasai, T. Matsuda, A. Tonomura, 
and V.V. Moshchalkov,
``Direct observation of vortex dynamics in superconducting films with 
regular arrays of defects.''
{\em Science} {\bf 274}, pp. 1167-1170, 1996.   

\bibitem{Commensurate} 
C. Reichhardt, C.J. Olson, and F. Nori,
``Commensurate and incommensurate vortex states in superconductors with
periodic pinning arrays.''
{\em Phys.~Rev.~B} {\bf 57}, pp.7937--7943, 1998.  

\bibitem{Jensen}
C.~Reichhardt and N. Gr{\o}nbech-Jensen,
``Collective multivortex states in periodic arrays of traps.''
{\em Phys.~Rev. Lett.} {\bf 85}, pp. 2372--2375, 2000. 

\bibitem{Scalettar}
C. Reichhardt, C.J. Olson, R.T.~Scalettar, 
and G.T.~Zim{\' a}nyi,  
``Commensurate and incommensurate vortex lattice melting in 
periodic pinning arrays.''
{\em Phys.~Rev.~B.} {\bf 64}, pp. 145409, 2001.

\bibitem{Korda}
P.T. Korda, M.B. Taylor, and D.G. Grier,
``Kinetically locked-in colloidal transport in an array of optical tweezers.''
{\em Phys.~Rev.~Lett.} {\bf 89}, pp. 128301/1--4, 2002.

\bibitem{Korda2}
P.T. Korda, G.C. Spalding, and D.G. Grier,
``Evolution of a colloidal critical state in an optical pinning potential 
landscape.''
{\em Phys.~Rev.~B} {\bf 66}, pp. 024504/1--7, 2002. 

\bibitem{Spalding}
M.P. MacDonald, G.C. Spalding, and K.~Dholakia,  
``Microfluidic sorting in an optical lattice.''
{\em Nature} {\bf 426}, pp. 421--424, 2003.

\bibitem{Smith} 
R.L. Smith, G.C. Spalding, K. Dholakia, and M.P. MacDonald,  
``Colloidal sorting in dynamic optical lattices.''
{\em J. Optics A} {\bf 9}, pp. S134--S138, 2007.

\bibitem{MP} 
M. MacDonald, G. Spalding, and K.~Dholakia, 
``All-optical sorting.''
{\em Opt. Photon. News} {\bf 15}(12), pp. 23, 2004. 

\bibitem{Sm}
C. Reichhardt and C.J. Olson Reichhardt, 
``Dynamic regimes and spontaneous symmetry breaking for 
driven colloids on triangular substrates.''
{\em Europhys. Lett.} {\bf 68}, pp. 303--309, 2004.

\bibitem{Sancho}
A.M. Lacasta, J.M. Sancho, A.H. Romero, and K. Lindenberg,
``Sorting on periodic surfaces.''
{\em Phys. Rev. Lett.} {\bf 94}, pp. 160601, 2005.

\bibitem{Xiao}
K. Xiao and D.G. Grier,
``Multidimensional optical fractionation of colloidal 
particles with holographic verification.''
{\em Phys. Rev. Lett.} {\bf 104}, pp. 028302, 2010.

\bibitem{Sohn}
M. Balvin, E. Sohn, T. Iracki, G. Drazer, and J. Frechette,
``Directional locking and the role of irreversible 
interactions in deterministic hydrodynamics
separations in microfluidic devices.''
{\em Phys. Rev. Lett.} {\bf 103}, pp. 078301, 2009.

\bibitem{Speer}
D. Speer, R. Eichhorn, and P. Reimann,
``Exploiting lattice potentials for sorting chiral particles.''
{\em Phys. Rev. Lett.} {\bf 105}, pp. 090602, 2010.

\bibitem{Prost}
J. Rousselet, L. Salome, A. Ajdari, and J. Prost,
``Directional motion of Brownian particles induced by a 
periodic asymmetric potential.''
{\em Nature} {\bf 370}, pp. 446--448, 1994.

\bibitem{L}
L.P. Faucheux, L.S. Bourdieu, P.D. Kaplan, and A.J. Libchaber,
``Optical thermal ratchet.''
{\em Phys. Rev. Lett.} {\bf 74}, pp. 1504--1507, 1995.

\bibitem{P}
P. Reimann,
``Brownian motors: noisy transport far from equilibrium.''
{\em Phys. Rep.} {\bf 361}, pp. 1--265, 2002.

\bibitem{Ratchet}
C. Reichhardt, C.J. Olson, and M.B. Hastings,
``Rectification and phase locking for particles on symmetric 
two-dimensional periodic substrates.''
{\em Phys. Rev. Lett.} {\bf 89}, pp. 024101, 2002.

\bibitem{Ratchet2}
C. Reichhardt and C.J. Olson Reichhardt,  
``Absolute transverse mobility and ratchet effect on periodic two-dimensional
symmetric substrates.''
{\em Phys. Rev. E} {\bf 68}, pp. 046102, 2003.

\bibitem{Tierno}
P. Tierno, T.H. Johansen, and T.M. Fischer,
``Localized and delocalized motion of colloidal particles on
a magnetic bubble lattice.''
{\em Phys. Rev. Lett.} {\bf 99}, pp. 038303, 2007.

\bibitem{Hastings}
M.B. Hastings, C.J. Olson Reichhardt, and C. Reichhardt,
``Ratchet cellular automata.''
{\em Phys. Rev. Lett.} {\bf 90}, pp. 247004, 2003.

\bibitem{Babic}
D. Babic and C. Bechinger,
``Noise-enhanced performance of ratchet cellular automata.''
{\em Phys. Rev. Lett.} {\bf 94}, pp. 148303, 2005.

\bibitem{Lee}
S.-H. Lee, K. Ladavac, M. Polin, and D.G. Grier,
``Observation of flux reversal in a symmetric optical thermal
ratchet.''
{\em Phys. Rev. Lett.} {\bf 94}, pp. 110601, 2005.

\bibitem{Ren}
R. Gommers, S. Denisov, and F. Renzoni,
``Quasiperiodically driven ratchets for cold atoms.''
{\em Phys. Rev. Lett.} {\bf 96}, pp. 240604, 2006.

\bibitem{Lucas}
C. Mennerat-Robilliard, D. Lucas, S. Guibal, J. Tabosa, C. Jurczak,
J.Y. Courtois, and G. Grynberg,
``Ratchet for cold rubidium atoms: the asymmetric optical lattice.''
{\em Phys. Rev. Lett.} {\bf 82}, pp. 851--854, 1999.

\bibitem{Janko}
C.-S. Lee, B. Janko, I. Derenyi, and A.-L Barabasi,
``Reducing vortex density in superconductors using the `ratchet effect.' ''
{\em Nature} {\bf 400}, pp. 337--340, 1999.

\bibitem{Marchesoni}
J.F. Wambaugh, C. Reichhardt, C.J. Olson, F. Marchesoni, and
F. Nori,
``Superconducting fluxon pumps and lenses.''
{\em Phys. Rev. Lett.} {\bf 83}, pp. 5106--5109, 1999.

\bibitem{Olson2}
C.J. Olson, C. Reichhardt, B. Janko, and F. Nori,
``Collective interaction-driven ratchet for transporting flux quanta.''
{\em Phys. Rev. Lett.} {\bf 87}, pp. 177002, 2001.

\bibitem{Silva}
C.C. de Souza Silva, J. Van de Vondel, M. Morelle, and
V.V. Moshchalkov,
``Controlled multiple reversals of a ratchet effect.''
{\em Nature} {\bf 440}, pp. 651--654, 2006.

\bibitem{Lu}
Q.M. Lu, C.J. Olson Reichhardt, and C. Reichhardt,
``Reversible vortex ratchet effects and ordering in superconductors with
simple asymmetric potential arrays.''
{\em Phys. Rev. B} {\bf 75}, pp. 054502, 2007.

\bibitem{Farkas}
Z. Farkas, F. Szalai, D.E. Wolf, and T. Vicsek,
``Segregation of granular binary mixtures by a ratchet mechanism.''
{\em Phys. Rev. E} {\bf 65}, pp. 022301, 2002.

\bibitem{Wambaugh}
J.F. Wambaugh, C. Reichhardt, and C.J. Olson,
``Ratchet-induced segregation and transport of nonspherical grains.''
{\em Phys. Rev. E} {\bf 65}, pp. 031308, 2002.

\bibitem{Linke}
H. Linke, B.J. Aleman, L.D. Melling, M.J. Taormina, M.J. Francis,
C.C. Dow-Hygelund, V. Narayana, R.P. Taylor, and A. Stout,
``Self-propelled Leidenfrost droplets.''
{\em Phys. Rev. Lett.} {\bf 96}, pp. 154502, 2006.

\bibitem{Andras}
A. Lib{\' a}l, C. Reichhardt, B. Janko, and C.J. Olson Reichhardt,
``Dynamics, rectification, and fractionation for colloids on flashing 
substrates.''
{\em Phys. Rev. Lett.} {\bf 96}, pp. 188301, 2006.

\bibitem{Howse}
J.R. Howse, R.A.L. Jones, A.J. Ryan, T. Gough, R. Vafabakhsh, and 
R. Golestanian,
``Self-motile colloidal particles: From directed propulsion to 
random walk.''
{\em Phys. Rev. Lett.} {\bf 99}, pp. 048102, 2007.

\bibitem{Dreyfus}
R. Dreyfus, J. Baudry, M.L. Ropre, M. Fermigier, H.A. Stone, and J. Bibette,
``Microscopic artificial swimmers.''
{\em Nature} {\bf 437}, pp. 862--865, 2005.

\bibitem{Berg}
H.C. Berg, {\it Random Walks in Biology.}
Princeton University Press, Princeton, NJ, 1993.

\bibitem{Austin}
P. Galajda, J. Keymer, P. Chaikin, and R. Austin,
``A wall of funnels concentrates swimming bacteria.''
{\em J. Bacteriol.} {\bf 189}, pp. 8704--8707, 2007.

\bibitem{Wan}
M.B. Wan, C.J. Olson Reichhardt, Z. Nussinov, and C. Reichhardt, 
``Rectification of swimming bacteria and self-driven particle
systems by arrays of asymmetric barriers.''
{\em Phys. Rev. Lett.} {\bf 101}, pp. 018102, 2008.

\bibitem{Ruocco}
L. Angelani, R. Di Leonardo, and G. Ruocco,
``Self-starting micromotors in a bacterial bath.''
{\em Phys. Rev. Lett.} {\bf 102}, pp. 048104, 2009.

\bibitem{Ruocco2}
R. Di Leonardo, L. Angelani, D. Dell'Arciprete, G. Ruocco,
V. Iebba, S. Schippa, M.P. Conte, F. Mecarini, F. De Angelis,
and E. Di Fabrizio,
``Bacterial ratchet motors.''
{\em Proc. Nat. Acad. Sci. (USA)} {\bf 107}, pp. 9541--9545, 2010.

\bibitem{Aranson}
A. Solkolov, M.M. Apodaca, B.A. Grzybowski, and I.S. Aranson,
``Swimming bacteria power microscopic gears.''
{\em Proc. Nat. Acad. Sci. (USA)} {\bf 107}, pp. 969--974, 2010.

\bibitem{Tailleur}
J. Tailleur and M.E. Cates,
``Sedimentation, trapping, and rectification of dilute bacteria.''
{\em EPL} {\bf 86}, pp. 60002, 2009.

\bibitem{Peter}
P. Galajda, J. Keymer, J. Dalland, S. Park, S. Kou, and
R. Austin,
``Funnel ratchets in biology at low Reynolds number: choanotaxis.''
{\em J. Modern Optics} {\bf 55}, pp. 3413--3422, 2008.

\bibitem{Shear}
B. Kaehr and J.B. Shear,
``High-throughput design of microfluidics based on directed
bacterial motility.''
{\em Lab Chip} {\bf 9}, pp. 2632--2637, 2009.

\bibitem{Whitesides}
S.E. Hulme, W.R. DiLuzio, S.S. Shevkoplyas, L. Turner, M. Mayer,
H.C. Berg, and G.M. Whitesides,
``Using ratchets and sorters to fractionate motile cells of
Escherichia coli by length.''
{\em Lab Chip} {\bf 8}, pp. 1888--1895, 2008.

\bibitem{Mahmud}
G. Mahmud, C.J. Campbell, K.J.M. Bishop, Y.A. Komarova,
O. Chaga, S. Soh, S. Huda, K. Kandere-Grzybowska, and
B.A. Grzybowski,
``Directing cell motions on micropatterned ratchets.''
{\em Nature Phys.} {\bf 5}, pp. 606--612, 2009.

\end{thebibliography}
\end{document}